\documentclass[doublecol]{epl2}

\newcommand{\bn}{\begin{enumerate}}
\newcommand{\en}{\end{enumerate}}
\newcommand{\ba}{\begin{eqnarray}}
\newcommand{\ea}{\end{eqnarray}}

\usepackage{graphicx,color}

\newcommand{\be}{\begin{equation}}
\newcommand{\ee}{\end{equation}}

\newcommand{\et}{{\it et al. }}
\newcommand{\ete}{{\it et al.}}

\def\prl{{ Phys. Rev. Lett. }}

\def\prb{{ Phys. Rev. B }}

\title{ {Switching ferromagnetic spins by an ultrafast laser
  pulse:}\\ Emergence of giant optical spin-orbit torque
}
\shorttitle{Spin reversal} 

\author{G. P. Zhang\inst{1} \and Y. H. Bai\inst{2} \and Thomas F. George\inst{3}}
\shortauthor{G. P. Zhang \etal}

\institute{
  \inst{1}Department of Physics, Indiana State University,
   Terre Haute, IN 47809, USA\\
  \inst{2}Office of Information Technology, Indiana State
  University, Terre Haute, IN 47809, USA \\
\inst{3}Office of the Chancellor and Center for Nanoscience
  \\Departments of Chemistry \& Biochemistry and Physics \& Astronomy
  \\University of Missouri-St. Louis, St.  Louis, MO 63121, USA
}


\abstract{Faster magnetic recording technology is indispensable to
  massive data storage and big data sciences. {All-optical
    spin switching offers a possible solution}, but at present it is
  limited to a handful of expensive and complex rare-earth
  ferrimagnets.  The spin switching in more abundant ferromagnets may
  significantly expand the scope of all-optical spin switching.  Here
  by studying 40,000 ferromagnetic spins, we show that it is the
  optical spin-orbit torque that determines the course of spin
  switching in both ferromagnets and ferrimagnets.  Spin switching
  occurs only if the effective spin angular momentum of each
  constituent in an alloy exceeds a critical value. Because of the
  strong exchange coupling, the spin switches much faster in
  ferromagnets than weakly-coupled ferrimagnets. This establishes a
  paradigm for all-optical spin switching.  The resultant magnetic
  field (65 T) is so big that it will significantly reduce high
  current in spintronics, thus representing the beginning of
  photospintronics.  }

\begin{document}

\maketitle

\section{Introduction}

Magnetic switching is the single most important operation for any
modern magnetic storage device, where a magnetic field is employed to
switch microscopic spins from one direction to another. However, as
the areal density increases, the switching speed becomes a major
bottleneck for future technological advancement. A possible solution
emerged when Beaurepaire \et \cite{eric} reported that a 60-fs laser
pulse reduced the spin moment of ferromagnetic nickel films within 1
ps.  Their finding heralded the arrival of
femtomagnetism {\cite{ourreview,prl00,rasingreview}}, and
research efforts intensified immediately\cite{kimel,np09}. However,
for over a decade, the focus has been on demagnetization, not magnetic
switching. A major breakthrough came when Stanciu and coworkers
\cite{stanciu} demonstrated that a single laser pulse could
permanently switch the magnetic spin orientation in amorphous GdFeCo
samples.  This all-optical helicity-dependent spin switching (AOS)
ignited the research community since it may be an alternative to the
current magnetic storage technology\cite{rasingreview}. However, most
AOS samples are amorphous \cite{hass,mangin} and are hard to simulate
without significant approximations.  To this end, a unified
understanding is still missing, but several promising mechanisms have
been proposed, which include the inverse Faraday effect
\cite{stanciu,vahaplarprb}, spin-flip stimulated Raman scattering
\cite{gr,pop}, magnetic circular dichroism\cite{khorsand}, magnetic
sublattice competition\cite{mentink}, pure thermal effect
\cite{ostler,atxitia} and ultrafast exchange scattering\cite{bar}.
Recently, Lambert \et \cite{lambert} reported AOS in an ultrathin
ferromagnetic [Co(0.4 nm)/Pt(0.7 nm)]$_3$ multilayer.
{Medapalli \et \cite{med} demonstrated that the
  helicity-dependent switching in Co/Pt proceeds in two steps
  \cite{hadri}.}  Such a system is much more amenable to the
simulation without any major approximation, and its magnetic
properties have been well known for some time\cite{sod}. It is likely
that a detailed study of such a system may shed new light on AOS.


\section{Spin reversal theory}

\begin{figure}
\includegraphics[angle=0,width=8.5cm]{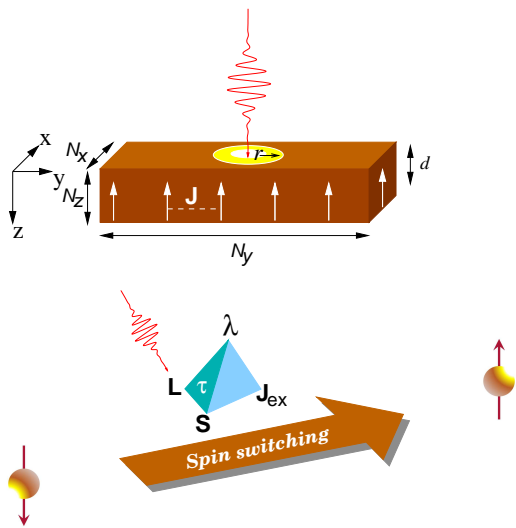}
\caption{ All-optical spin switching in ferromagnets.  (Top) The
  simulated sample has the dimensions $(N_x=101)\times (N_y=101)\times
  (N_z=4)$, more than 40,000 spins. The light propagates along the
  $+z$ direction with penetration depth $d$ and radius of the spot
  $r$.  (Bottom) The laser-induced optical spin-orbit torque provides
  the necessary torque to reverse the spin.  }
\label{fig1}
\end{figure}

We employ a thin film of $101\times 101\times 4$ or 40,804 lattice
sites in a simple cubic structure (see the top half of
Fig. \ref{fig1}) {with an open boundary condition}.  Each
site has a spin ${\bf S}_i$ which is exchange-coupled to the nearest
neighboring spins through the exchange interaction $J_{ex}$.  {Our
Hamiltonian \cite{jpcm11,jpcm13,epl15,mplb16}}, which is
often used in magnetic multilayers\cite{hs}, is \ba H&=&\sum_i \left
[\frac{{\bf p}_i^2}{2m}+V({\bf r}_i) +\lambda {\bf L}_i\cdot {\bf S}_i
  -e {\bf E}({\bf r}, t) \cdot {\bf r}_i\right
]\nonumber\\ &-&\sum_{ij}J_{ex}{\bf S}_i\cdot {\bf S}_{j}, \label{ham}
\ea where the first term is the kinetic energy operator of the
electron, the second term is the potential energy operator, $\lambda$
is the spin-orbit coupling in units of eV/$\hbar^2$, $ {\bf L}_i$ and
$ {\bf S}_i $ are the orbital and spin angular momenta at site $i$ in
the unit of $\hbar$, respectively, and {\bf p} and {\bf r} are the
momentum and position operators of the electron, respectively.  To
minimize the number of parameters, we choose a spherical harmonic
potential $V({\bf r}_i)=\frac{1}{2}m\Omega^2 {\bf r}_i^2$ with system
frequency $\Omega$, but this approximation can be lifted when accurate
potentials are known.  {Our model represents a small step
  towards a complete model}.
We assume that the electron moves along the $z$
axis with an initial velocity of 1 nm/fs in the harmonic potential, so
the initial orbital angular momentum is zero.  The last term is the
exchange interaction, and $J_{ex}$ is the exchange integral in units
of eV/$\hbar^2$. Although our main interest is in ferromagnets, the
same Hamiltonian can describe both antiferromagnets and ferrimagnets.
Such a Hamiltonian contains the necessary ingredients for AOS.

Figure \ref{fig1} shows that a laser pulse propagates along the $+z$
axis; its amplitude is attenuated according to Beer's law ${e}^{-z/d}$
(along $+z$), where $d$ is the penetration depth.  The bottom half of
Fig. \ref{fig1} illustrates our idea of spin torque to switch spins.
For convenience, the spatial dimension is measured in the unit of the
lattice site number along each direction, so that all the spatial
variables are dimensionless or in the unit of the site number. The
laser spot is centered at $x_c=51$ and $y_c=51$ with radius $r$ and
lateral spatial profile\cite{vahaplarprb}
$e^{-[(x-x_c)^2+(y-y_c)]^2/r^2}$ (in the $xy$ plane).  The laser
electric field is described by \be {\bf E}({\bf r},t)={\bf
  A}(t)\exp[-\frac{(x-x_c)^2+(y-y_c)^2}{r^2}-\frac{z}{d}], \ee where
$x$ and $y$ are the coordinates in the unit of the site number.  Since
in the following our spins are all initialized along the $-z$ axis, we
choose a left-circularly polarized field ${\bf A}(t)$ which has a
Gaussian shape $ {\bf A}(t)=A_0{\rm e}^{-t^2/T^2}[- \sin(\omega t)
  \hat{x}+\cos(\omega t) \hat{y} ], $ where $\omega$ is the laser
carrier frequency, $T$ is the laser pulse duration, $A_0$ is the laser
field amplitude, $t$ is time, $\hat{x}$ and $\hat{y}$ are unit
vectors, respectively. We choose $T=100$ fs.  {We} only consider a
resonant excitation where the laser photon energy $\hbar\omega=1.6$ eV
matches the system energy $\hbar\Omega$; for an off-resonant
excitation, we refer the reader to a prior study\cite{epl15}. In
transition metals, the penetration depth is about 14 nm, which
corresponds to 30 layers, so we choose $d=30$. To compute the spin
evolution, we employ the Heisenberg's equation of motion,
$i\hbar\dot{A}=[A,H]$, {where we make the time-dependent Hartree-Fock
  approximation, so that all the operators are replaced by their
  respective expectation values}, and then we solve the equation
numerically. {Our calculation of the spin change is similar
  to that of Wienholdt \et \cite{wienholdt} though they used a thermal
  field. }

\section{Dependence of spin switching on spin angular momentum}

\begin{figure}
\includegraphics[angle=0,width=8.5cm]{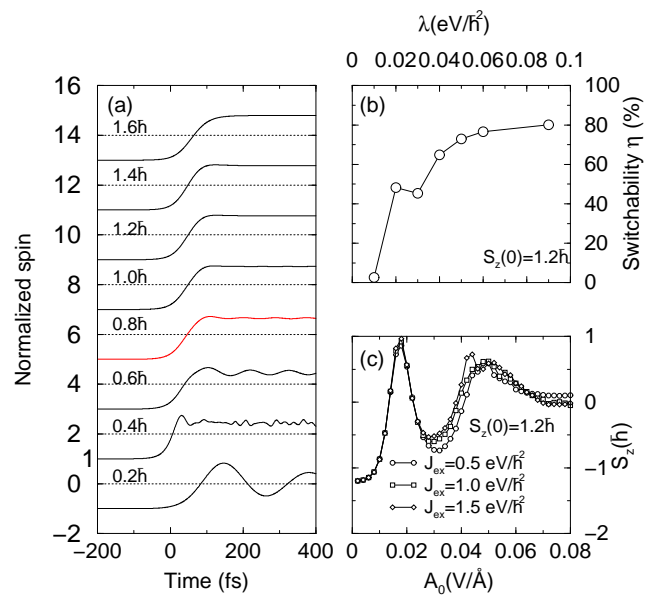}
\caption{ (a) Time evolution of the $z$ component of the normalized
  and system-averaged spin angular momentum for eight initial spin
  values $S_z(0)$ from 0.2$\hbar$ to 1.6$\hbar$.  The spin reversal
  realized starts once $S_z(0)$ is around and above 0.8$\hbar$.  (b)
  Switchability as a function of spin-orbit coupling. The critical
  value is around 0.04 eV/$\hbar^2$. (c) Dependence of the final spin
  on the laser field amplitude for three values of the exchange
  integral $J_{ex}$.  }
\label{fig20}
\end{figure}

We choose eight initial spin momenta $S_z(0)$ from $0.2\hbar$ to
1.6$\hbar$ in steps of 0.2$\hbar$, which covers most magnetic
materials.  For each $S_z(0)$, we vary the laser field
amplitude {\cite{prl00,med}} $A_0$ from 0.01 to 0.08 $\rm V/\AA$ in steps
of 0.002 $\rm V/\AA$. This step is tedious but necessary, since
different $S_z(0)$ have different optimal field amplitudes for spin
reversal.  We fix the spin-orbit coupling at $\lambda=0.06 {\rm
  eV}/\hbar^2$, the exchange interaction $J_{ex}$ at 1 ${\rm
  eV}/\hbar^2$, and the spot radius of $r=100$. { The spins are
  initialized along the $-z$ axis, equivalent to applying a magnetic
  uniaxial anisotropy.}  A spin reversal is considered achieved if the
$z$ component spin angular momentum $S_z$ changes from a negative
value to a large and positive value at the end of the dynamics. Figure
\ref{fig20}(a) shows the normalized and system-averaged spin as a
function of time for each $S_z(0)$ at its respective optimal laser
field amplitude.  All the curves, except $S_z(0)=0.2\hbar$, are
vertically shifted for clarity.  The dotted horizontal lines denote
$0\hbar$.  We start with $S_z(0)=0.2\hbar$, and we see that the spin
does not switch and only oscillates around $0\hbar$ with a period
determined by the product of $\lambda$ and
$S_z(0)$\cite{epl15,jap15}. When we increase $S_z(0)$ to 0.4$\hbar$,
the oscillation is attenuated and the final spin is barely above
$0\hbar$. And the situation does not change much for
$S_z(0)=0.6\hbar$. However, when we continue to increase $S_z(0)$
above $0.8\hbar$, the spin ringing is strongly reduced, and the final
spin settles down at a large positive value, an indication of spin
reversal. Above $0.8\hbar$, the situation gets better. For this
reason, we define a critical spin angular momentum $S_z^c=0.8\pm 0.2
\hbar$ for AOS.

{ To quantify AOS, we define the spin switchability as $
  \eta=\frac{S_z^f}{S_z(0)}\times 100\%,$ where $S_z^f$ is the final
  spin angular momentum. This definition is different from that of
  Vahaplar \ete\cite{vahaplarprb}}. We fix $S_z(0)=1.2\hbar$, but
change the spin-orbit coupling $\lambda$. Note that our conclusions
are the same for different $S_z(0)$ as far as it is above $S_z^c$.
Figure \ref{fig20}(b) shows that a minimum $\lambda$ of 0.04
eV$/\hbar^2$ is required to reverse spins.  Too small a $\lambda$ only
leads to a strong spin oscillation, regardless of the laser field
amplitude. This indicates a unique role of spin-orbit coupling (SOC)
in AOS. The roles of the exchange interaction and laser field
amplitude are shown in Fig. \ref{fig20}(c), where we fix
$S_z(0)=1.2\hbar$, $r=100$ and $\lambda=0.06 \rm~eV/\AA^2$. We notice
that as $A_0$ increases, $S_z$ sharply increases and reaches its
maximum. If we increase it further, $S_z$ is reduced since the spin
overshoots, and an asymmetric peak is formed. This constitutes our
first criterion that the laser amplitude must fall into a narrow
region for AOS to occur.  {This is consistent with Medapalli
  \ete's finding (see Fig. 1(c) of their paper \cite{med}); such a
  helicity-dependent switching also agrees with another study by El
  Hadri \et \cite{hadri}}. {These agreements do not
  necessarily validate all the aspects of our model but instead they
  simply suggest that our model may offer an alternative to the
  existing models.}
 If we increase $A_0$ further, a second
peak appears since the spin re-switching starts.  These double peaks
do not appear for a smaller $S_z(0)$.  We find that the exchange does
not change this dependence a lot.

\section{Phase diagram of spin reversal}

\begin{figure}
\includegraphics[angle=0,width=8.5cm]{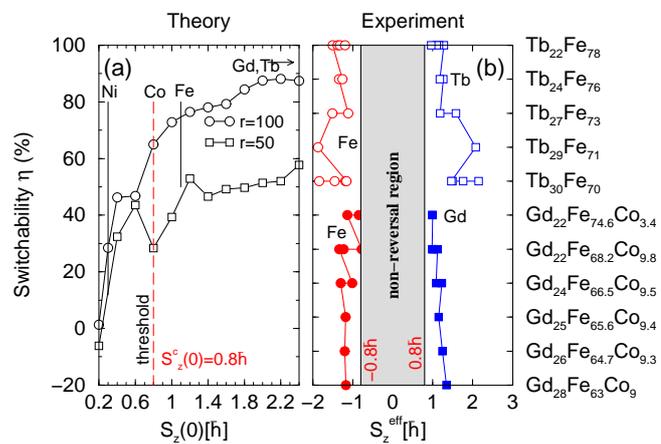}
\caption{ (a) Phase diagram of the spin switchability versus the
  initial spin angular momentum $S_z(0)$ at the respective optimal
  laser field amplitudes. The empty circles and boxes refer to the
  results with $r=100$ and $r=50$, respectively. The long-dashed line
  denotes the critical spin $S_z^c$.  Two thin vertical lines
  represent the spins for Ni and Fe. Co is on the border line, while
  Gd and Tb are way above $S_z^c$. (b) Computed experimental effective
  spin angular momentum for each element in 11 GdFeCo and TbFe
  alloys\cite{hassdenteufel2015}. Without exception, all elements have
  spin larger than $S_z^c$.}
\label{fig3}
\end{figure}

We construct a phase diagram of spin reversal ($\eta-S_z(0)$) in
Fig. \ref{fig3}(a) for thirteen $S_z(0)$'s and two radii of the laser
spot, $r=100$ and 50.  For $\eta$ to exceed 50-60\%, $S_z(0)$ must be
higher than the critical value of $S_z^c=0.8\pm 0.2\hbar$.  The
long-dashed line denotes $S_z^c$. We see that the nickel's spin
momentum is well below $S_z^c$, which explains why nickel has never
been used for AOS.  Co is on the threshold.  In Co-Pt granular
samples\cite{figueroa}, the effective spin magnetic moment per $3d$
hole is 0.77 $\mu_{\rm B}$; since there are 2.49-2.62 holes, the spin
angular momentum is 0.96$\hbar$, satisfying this criterion.  In the
ultrathin ferromagnetic [Co(0.4 nm)/Pt(0.7 nm)]$_3$
films\cite{lambert}, due to the reduced dimensionality, the enhanced
spin moment greatly increases the chance for AOS.  The empty boxes in
Fig. \ref{fig3}(a) represent the case with $r=50$ (which is close to
the switch limit), where only a small portion of the sample is exposed
to the laser light. We see that the switchability reduces sharply
since the laser fluence on lattice sites away from the center of the
laser beam becomes very weak and is not strong enough to reverse spins
on those sites.  {Since the essence of AOS is rooted in
  spin-orbit coupling and all the switchabilities are obtained at the
  optimal field amplitude, we do not expect that a more accurate
  potential would change the phase diagram strongly.}  Our criterion
not only applies to ferromagnets, but also to ferrimagnets. Figure
\ref{fig3}(b) illustrates that each of the major elements in all the
11 GdFeCo and TbFe alloys\cite{hassdenteufel2015} has the effective
spin above $S_z^c$. This constitutes strong evidence that our finding
has a broader impact on the ongoing research in all-optical spin
switching.

\section{Emergence of optical spin-orbit torque}

\begin{figure}
\includegraphics[angle=0,width=8cm]{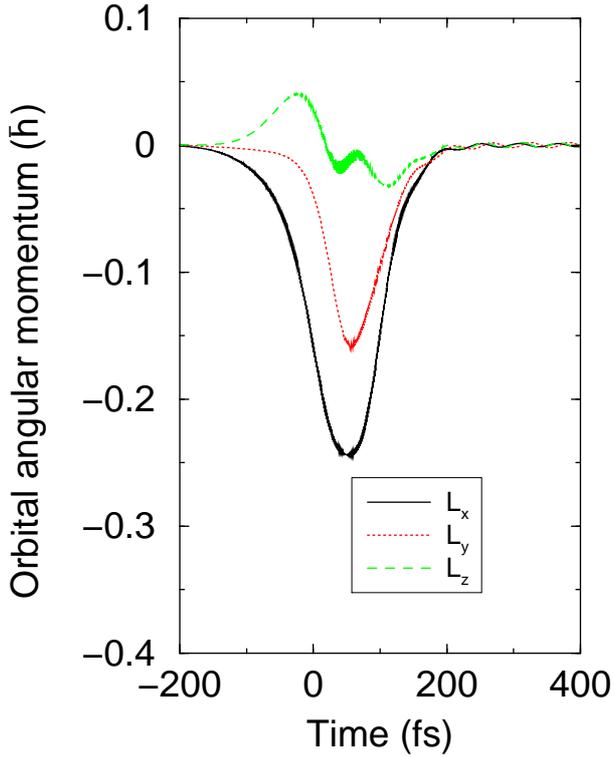}
\caption{Orbital angular momentum change as a function of time.  Here
  $J_{ex}=1{\rm eV}/\hbar^2$, and $S_z(0)=1.2\hbar$. The laser
  amplitude is at its optimal value of 0.018$\rm V/\AA$.  The solid,
  dotted and dashed lines denote $L_x$, $L_y$ and $L_z$ components
  {\cite{mplb16}}, respectively.  }
\label{orbital}
\end{figure}

While the effect of the laser field amplitude on AOS is obvious
\cite{vahaplarprl}, how the initial spin $S_z(0)$ affects the spin
switching is not obvious.  We examine how the spin evolves with time.
For a spin at site $i$, the spin angular momentum ${\bf S}$ precedes
according to \be \frac{d{\bf S}_i}{dt}= \sum_{j(i)} J_{ex} {\bf
  S}_i\times {\bf S}_j +\lambda ({\bf L}_i \times {\bf S}_i),
 \label{s} \ee where the two driving terms on the right-hand
side represent two torques.  The first is the Heisenberg exchange
torque $\tau_{ex}=\sum_{j(i)} J_{ex} {\bf S}_i\times {\bf S}_j$. Since
all the spins are ferromagnetically ordered, this torque is very
small. The second one is the spin-orbit torque (SOT),
$\tau_{soc}=\lambda ({\bf L}_i \times {\bf S}_i)$, {which may
  serve as a source term for the inverse Faraday effect
  \cite{john,berritta}.}  Before the laser excitation, $\tau_{soc}$ is
small, since in solids the orbital angular momentum ${\bf L}$ is
largely quenched. With the arrival of the laser pulse, ${\bf L}$ is
boosted sharply {\cite{john}} (see Fig. \ref{orbital}) and
helicity-dependent, where $J_{ex}=1{\rm eV}/\hbar^2$, and
$S_z(0)=1.2\hbar$, but three components of the orbital angular
momentum behave differently.  $L_x$ and $L_y$ are mostly negative, but
$L_z$ is positive. Around 50 fs, $L_x$ reaches $-0.24\hbar$, while
$L_y$ swings to $-0.16\hbar$ and the change in $L_z$ is smaller,
around $0.04\hbar$.  All three components settle down to zero around
200 fs. This is very important, since if the orbital momentum were big
after the laser field is gone, the spin would oscillate very strongly
and could not be reversed faithfully.  Thus, through the spin-orbit
coupling, the laser field increases the orbital angular momentum, and
subsequently $\tau_{soc}$ is boosted. For this reason, Tesavova
\et\cite{tesarova} called $\tau_{soc}$ the optical spin-orbit torque,
or femtosecond spin-orbit torque by Lingos \et \cite{lingos}.

\begin{figure}
\includegraphics[angle=0,width=8.5cm]{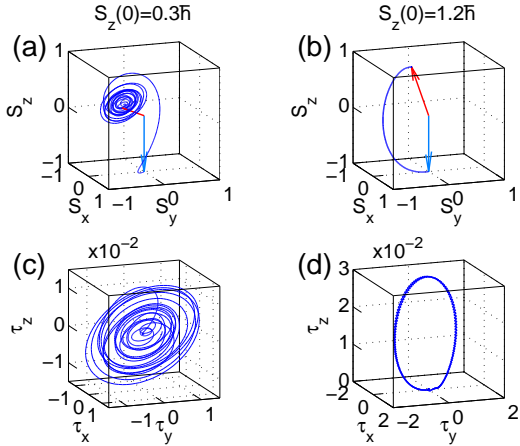}
\caption{ (a) Precession of the normalized spin angular momentum in
  3-dimensional space at $S_z(0)=0.3\hbar$. The blue arrow denotes the
  initial spin, and the red one the final spin. The trace of the final
  spin forms a spiral and the spin does not switch. (b) The normalized
  spin angular momentum at $S_z(0)=1.2\hbar$ is directly switched from
  the $-z$ to $+z$ axis without precession
  {\cite{mplb16}}. (c) Time evolution of the spin-orbit
  torque at $S_z(0)=0.3\hbar$.  The torque is zero in the beginning.
  All the torques are in the units of $\hbar$/fs.  (d) Same as (c) but
  for $S_z(0)=1.2\hbar$  {\cite{mplb16}}.}
\label{fig4}
\end{figure}

We choose two initial spin momenta, $S_z(0)=0.3\hbar$ and 1.2$\hbar$,
with all the spins initialized along the $-z$ axis (see the light blue
arrows in Figs. \ref{fig4}(a) and (b)). Figure \ref{fig4}(a) shows
that at $0.3\hbar$ the spin undergoes strong oscillations and shows
many spirals, but does not settle down to the $+z$ axis after the
laser pulse is gone (see the red arrow).  By contrast, at $1.2\hbar$
the spin flips over from the $-z$ to $+z$ axis within 110 fs, without
strong oscillation (see the solid red arrow).  To understand why the
initial spin angular momentum has such a strong effect on AOS, Figure
\ref{fig4}(c) shows that $\tau_{soc}$ at $0.3\hbar$ is very weak,
around 0.01 $\hbar$/fs, and more importantly, it rapidly swings
between positive and negative values, both of which are detrimental to
the spin reversal.  At $S_z(0)=1.2\hbar$, $\tau_{soc}$ is positive and
large, which allows the spin to switch over successfully.  This
suggests that SOT offers an alternative path to AOS (see the bottom
figure of Fig. \ref{fig1}), and it acts like an effective magnetic
field, which has been sought after in the literature
\cite{vahaplarprb,ostler} for nearly a decade.  At 1.2$\hbar$, we
time-integrate the torque from -200 to +200 fs and find that the
time-averaged torque corresponds to 65 T of a magnetic field. In
spintronics, the spin transfer torque heavily relies on the high
electric current\cite{hs,stiles2}. Such a large SOT, if implemented in
real experiments, should significantly reduce the requirement of huge
electric current for spintronics\cite{bokor}, and thus opens a door
for rapid applications in storage technology\cite{wolf}.

\section{Conclusion}

We have investigated all-optical spin switching in
40,000 ferromagnetic spins. We identify that it is the laser-induced
optical spin-orbit torque that determines the fate of spin switching.
The spin-orbit torque sensitively depends on the value of the initial
spin momentum of each active element in a sample, regardless of the
types of magnets. To switch, each active element must have its
effective spin angular momentum larger than $(0.8\pm
0.2)\hbar$. { This means that the switchability in Fe, Gd
  and Tb is likely to be higher than Co and Ni. PMA observed in
  various AOS materials \cite{mangin} seems to be an indication of
  enhanced spin moment, which is in line with our theory. } The ps
all-optical spin switching observed in ferrimagnets is associated with
the weak exchange coupling; in ferromagnets, with a stronger coupling,
the switching is much faster.  SOT is so large that it will
significantly reduce the electric current used in spintronics.  After
our present study was finished, we noticed a recent publication by
Bokor's group \cite{bokor} to use a laser to assist magnetization
reversal. A combination of photonics and spintronics represents the
arrival of photospintronics\cite{mondal}.

\acknowledgments We would like to thank Dr.  Hassdenteufel for sending
us the experimental results\cite{hassdenteufel2015}.  This work was
solely supported by the U.S. Department of Energy under Contract
No. DE-FG02-06ER46304. Part of the work was done on Indiana State
University's quantum cluster and high-performance computers.  The
research used resources of the National Energy Research Scientific
Computing Center, which is supported by the Office of Science of the
U.S. Department of Energy under Contract No. DE-AC02-05CH11231. This
work was performed, in part, at the Center for Integrated
Nanotechnologies, an Office of Science User Facility operated for the
U.S. Department of Energy (DOE) Office of Science by Los Alamos
National Laboratory (Contract DE-AC52-06NA25396) and Sandia National
Laboratories (Contract DE-AC04-94AL85000).

\newpage

\clearpage


{\bf Supplementary Materials}


{A main difference between ferrimagnets and ferromagnets
  is that ferrimagnets have magnetic sublattices while ferromagnets do
  not. To directly apply our results to ferrimagnets, we need to
  understand whether these magnetic sublattice spins behave similarly
  to those spins in ferromagnets.  Fortunately, we find that according
  to our simulation, at least in the weak laser field limit,
  left(right)-circularly polarized light only switches spin from
  down(up) to up(down), not the other way around. Therefore, flipping
  a sublattice spin in ferrimagnets is equivalent to flipping a spin
  in ferromagnets. This is the theoretical basis to apply our theory
  to ferrimagnets.}

In the following, we explain how the spin angular momentum is computed
from the experimental data.  Experimentally, the measured magnetic
property is often the remanent magnetization $m_{\rm R}$ in units of
$10^3$A/m, or equivalently emu/cc \cite{hassdenteufel2015}. To convert
magnetization to spin angular momentum, we need the volume of the
sample, but experimentally the volume of the sample is not given. This
makes a quantitative comparison nearly impossible.

We find a simple and powerful method which does not rely on the
experimental volume.  Take R$_x$T$_{1-x}$ alloy as an example, where R
stands for Tb or Gd and T stands for Fe. We ignore Co since its
concentration is too low.  We first compute the effective volume \be
V_{eff}=xV_{\rm R}+(1-x)V_{\rm T}, \ee where $V_{\rm R}$ and $V_{\rm
  T}$ are the supercell volume of pure element R and T.  Tb has a hcp
structure, with the lattice constants $a=3.601\rm \AA$ and
$c=5.6936\rm \AA$; Fe has a bcc structure with $a=2.8665\rm \AA$.
Then we multiply $m_{\rm R}$ by $V_{eff}$ to get the effective spin
moment for the alloy, i.e., $M_{eff}= m_RV_{eff}$.  Since $M_{eff}$ is
in the units of [Am$^2$], we convert it to the Bohr magneton $\mu_B$,
with the conversion factor of $0.10783 \times 10^{-3}$.

Szpunar and Kozarzewski \cite{sz} carried out extensive
calculations on transition-metal and rare-earth intermetallic compounds
by comparing their results with the experimental ones, and concluded
that it is reasonable to assume that the average magnetic moments of
the transition metals and of the rare earth metals are roughly
independent of structures. Then, the effective spin moment $M_{eff}$
can be approximately written as \be M_{eff}=xM_{\rm R}+(1-x)M_{\rm
  T}\equiv M^{eff}_{\rm R} + M^{eff}_{\rm T}, \ee where $M_{\rm R}$
and $M_{\rm T}$ are the spin moments of pure R and T, respectively.
Here the last equation defines the effective spin moment for R and T.

However, this single equation is not enough to compute $M_{\rm R}$ and
$M_{\rm T}$ since there are two unknowns for a single equation.  The
trick is that we use two sets of compositions, $x_1$ and $x_2$, so we
have two equations, \ba M_{eff}^{(1)}=x_1M_{\rm R}+(1-x_1)M_{\rm
  T}\\ M_{eff}^{(2)}=x_2M_{\rm R}+(1-x_2)M_{\rm T}, \ea where
$M_{eff}^{(1)}=m_R^{(1)}V_{eff}^{(1)}$ and
$M_{eff}^{(2)}=M_R^{(2)}V_{eff}^{(2)}$.  Here again we rely on the
assumptions that $M_{\rm R}$ and $M_{\rm T}$ do not change much with
composition change from $x_1$ to $x_2$.  When we choose $x_1$ and
$x_2$, we are always careful whether $M_{\rm R}$ or $M_{\rm T}$
changes sign, since experimentally the reported values are the
absolute value. In addition, it is always better to choose those $x_1$
and $x_2$ which have the same sign of $M_{\rm R}$ and $M_{\rm
  T}$. Choosing several different pairs of $(x_1,x_2)$ is crucial to a
reliable result.  Solving the above two equations, we can find $M_{\rm
  R}$ and $M_{\rm T}$.

Before we compute the spin angular momentum, we check whether the
computed spin moments $M_{\rm R}$ and $M_{\rm T}$ (in the units of
$\mu_B$) are within the respective value of each pure element, i.e.,
$M_{\rm Gd}^{\circ}=7.63 \mu_B$, $M_{\rm Tb}^{\circ}=9.34 \mu_B$, and
$M_{\rm Fe}^{\circ}=2.2 \mu_B$.  If the computed spin moment ($M_{\rm
  R}$ and $M_{\rm T}$) is far off from those spin moments, this
indicates that either our method or the experimental result is not
reliable; as a result, their spin angular momentum is not included in
our figure, but is included here. Once the spin moment passes this
test, we proceed to convert the spin moment to spin angular momentum.


Our method works better for Gd alloys than Tb alloys, since the former
has zero orbital angular momentum but the latter has a nonzero orbital
angular momentum.  For Gd and Fe, the orbital momentum is largely
quenched.  Assuming that the Lande $g$-factor is 2, we divide the spin
moments $M_{\rm R}$ and $M_{\rm T}$ by 2 to get the spin angular
momentum $S_{\rm R}$ and $S_{\rm T}$ in the unit of $\hbar$. To get
the effective spin angular momentum, we multiply $S_{\rm R}$ and
$S_{\rm T}$ with $x$ and $1-x$, respectively, i.e.,
\ba S_{\rm R}^{eff}&=&x
S_{\rm R}\\
S_{\rm T}^{eff}&=&(1-x) S_{\rm T}.
\ea  It is these two
effective spin angular momenta that we apply our above criterion to.
For Tb, our results have an uncertainty since its orbital angular
momentum in its alloys is unknown, although its orbital angular
momentum in pure Tb metal is 3.03$\hbar$.  Table \ref{tab0} shows the
orbital-free spin angular momentum for 11 alloys, where we adopt a
simple cubic structure for Fe since it matches the experimental values
better. These data are used to plot Fig. 3(b) of the main paper.

\begin{figure}
\includegraphics[angle=270,width=14cm]{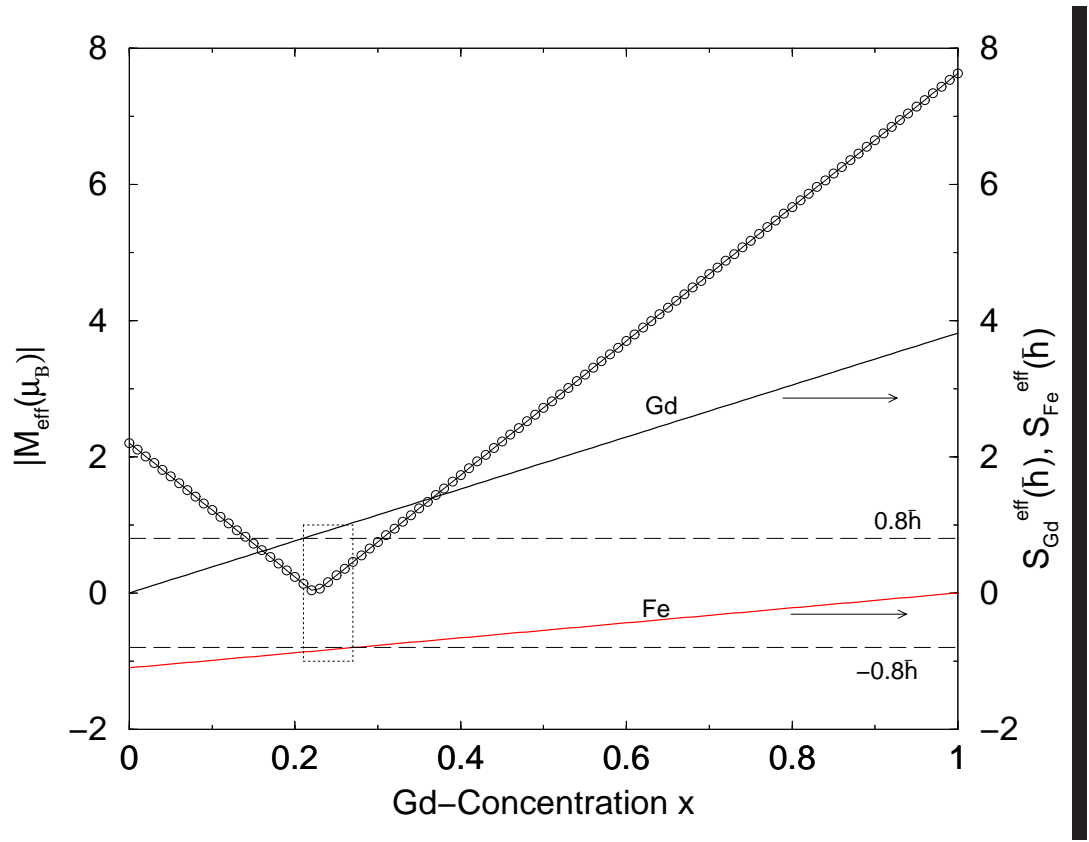}
\caption{ Effective spin moment change as a function of Gd
  concentration $x$ (circles). The two solid lines represent the effective
  spin angular momentum for Gd and Fe (using the right axis). The two
  horizontal dashed lines denote the predicted critical spin angular
  momentum ($\pm 0.8\hbar$). The dotted line box highlights the narrow
  region of the Gd concentration where spin angular momentum satisfies
  our criterion and the spin reversal occurs.  }
\label{supfig0}
\end{figure}

Before we show all the details of our results, we wish to present an
example how the spin moment changes with the concentration under our
assumption.  Since R and T are ferrimagnetically coupled and $M_{\rm
  R}$ and $M_{\rm T}$ differ by a sign, $M_{eff}$ changes from a
positive value to a negative as the composition $x$ changes.  Figure
\ref{supfig0} shows such an example, where we use the experimental value
$M_{\rm Gd}=7.63\mu_B$ of pure Gd and $M_{\rm Fe}=2.2\mu_B$ of pure
Fe. A V shape curve is formed, the same as the experiment
\cite{hassdenteufel2015}.  $ M_{eff}$ is close to zero around
$x=0.22$.  The effective spin momenta $S^{\rm eff}_{\rm Gd}$ for Gd
and $S^{\rm eff}_{\rm Fe}$ for Fe are also shown (use the right
axis). As $x$ increases, $S^{\rm eff}_{\rm Gd}$ increases but $|S^{\rm
  eff}_{\rm Fe}|$ is reduced. Two long-dashed lines denote our
critical values $\pm S_z^c$. We use the dotted line box to bracket the
narrow window for spin switching. Since this window is very close to
the compensation point (in term of the concentration $x$), this
explains why Hassdenteufel \et \cite{hass} found that the low remnant
magnetization for AOS must be below 125 emu/cc. This is the direct
consequence of the requirement of the critical value of $ S_z^c$.

In the following, we tabulate all the computed results for both GdFeCo
and TbFe alloys, respectively.  All the tables start with the spin
moment for each element, followed by the effective spin angular
momentum for each element in the alloys. To reduce possible errors in
those experimental data, we always choose multiple pairs of data
for the same alloy.

\subsection{Gd alloys}

We start with a pair of Gd$_{24}$Fe$_{66.5}$Co$_{9.5}$ and
Gd$_{22}$Fe$_{68.2}$Co$_{9.8}$. Table \ref{tab1} (the first two rows)
shows that Gd has a magnetic moment of -10.2187$\mu_B$ and Fe 3.9149
$\mu_B$, respectively, where we purposely keep more significant
figures to show the accuracy of our results.  Note that Gd and Fe are
ferrimagnetically coupled, so they differ by a negative sign.  By
comparing them with their respective element values, we conclude that
these moments are reasonable. We then compute the effective spin
angular momentum for Gd and Fe (see the third and fourth
columns). Clearly, both numbers are larger than our critical spin
angular momentum. Then we compute four additional combinations of
alloys. If the experimental results were exact and free of any error,
the obtained effective spin angular momentum should not
change. However, in reality, they do change, but we find that the
change for Gd alloys is very small. For instance,
Gd$_{22}$Fe$_{68.2}$Co$_{9.8}$ is used twice, but each case has a
similar $S^{eff}_{\rm Gd}$ and $S^{eff}_{\rm Fe}$ (compare the first
pair and third pair). The same is also true for
Gd$_{22}$Fe$_{74.6}$Co$_{3.4}$, but when we pair
Gd$_{22}$Fe$_{68.2}$Co$_{9.8}$ with Gd$_{22}$Fe$_{74.6}$Co$_{3.4}$, we
find a slightly larger change. The reason is easy to understand since
these two compounds have a very similar composition, and the relative
error becomes larger.

Table \ref{tab2} assumes a bcc structure for Fe. Here the values are
all reduced somewhat, but the main conclusion remains the same. We
also find that the biggest error comes from the alloy pairs with a
similar composition (see the last pair). We notice that $S^{eff}_{\rm
  Fe}$ is slightly below our critical value. It is likely that the
larger relative error when two compositions are close is responsible
for the discrepancy.

\subsection{Tb alloys}

In comparison with Gd alloys, Tb alloys are more prone to errors,
since their orbital momentum is not completely quenched. Table
\ref{tab3} shows two different stories. For the first seven pairs, we
see that all the moments for Tb and Fe are reasonably close to their
respective element moments. But for the last two pairs, their values
are way too low. We know why this occurs. Tb$_{36}$Fe$_{64}$ is not
the AOS compound, and only shows the pure thermal
demagnetization. From the first seven pairs, we see that the
experimental result for Tb$_{30}$Fe$_{70}$ is reliable, since
different pairs give a similar spin moment. But when it is paired with
Tb$_{36}$Fe$_{64}$, it leads to an unreasonable result. This means
that the structure-property of Tb$_{36}$Fe$_{64}$ is quite different
from the AOS compounds such as Tb$_{30}$Fe$_{70}$, and it may not have
the linear relation between the spin moment and the composition $x$ as
we assume above. Our finding is backed by the last pair, where
Tb$_{34}$Fe$_{66}$ is not an AOS compound initially, and only after
the heating does it become AOS. If we look at Hassdenteufel's
Fig. 7\cite{hassdenteufel2015}, we find that Tb$_{34}$Fe$_{66}$ does
not follow the trend of the rest of the TbFe alloys. For this reason,
they are not included in Table \ref{tab0}.  Table \ref{tab3} shows all
the spin angular moments that are computed, with zero orbital angular
momentum. Table \ref{tab4} shows the same data but with bcc structure
for Fe. The main conclusion is the same as Table \ref{tab3}.

\section{Helicity-dependent and helicity-independent all-optical spin
switchings}

There is enormous interest in both the all-optical helicity-dependent
spin switching (AO-HDS) and the all-optical helicity-independent spin
switching (AO-HIDS). Experimentally, Stanicu \et \cite{stanciu} first
demonstrated a clear helicity-dependent switching in GdFeCo, but when
they \cite{ostler} later increased the laser intensity, the switching
became helicity-independent. In other words, there is a clear
transition from a helicity-dependent switching to a
helicity-independent switching in GdFeCo when one increases the laser
fluence.  The underlying reason of this transition has been unclear,
though there are several mechanisms proposed \cite{ostler}. On the
other hand, Lambert \et \cite{lambert} demonstrated a clear
helicity-dependent switching in their $\rm Co(0.4~ nm)/Pt(0.7~ nm)]_3$
  film. At low laser power (362 nW), they showed that a reversed
  domain is written for $\sigma^+$, but not for $\sigma^-$ or linearly
  polarized light. When they increased the laser power, regions of
  demagnetized random domains developed. Therefore, the
  helicity-independent switching does not occur in CoPt films. This
  may suggest that the difference between AO-HDS and AO-HIDS can both
  be laser-intensity dependent and material-dependent. Our model,
  which is solely based on a ferromagnet, does show a
  helicity-dependent switching. The light helicity is important for
  our model to work, as far as the laser fluence is small. For
  instance, if the spin points down (the $-z$ axis), only the
  left-circularly polarized light can efficiently switch the spin up,
  if the laser field amplitude is weak. This finding appears to agree
  with the experimental results reasonably well \cite{lambert}. If the
  laser field becomes too stronger, we are not completely confident
  whether our model can describe the physics correctly, though we did
  test the model in a single site case \cite{epl15}, where we found
  that the switching becomes highly nonlinear, and even the linearly
  polarized light can switch the spin. For this reason, our present
  paper exclusively focuses on the lower laser field limit and
  ferromagnets. We are currently exploring whether our model can
  describe GdFeCo. Our present model, without further change, is
  unsuitable for GdFeCo since GdFeCo is amorphous, ferrimagnetic and
  much more complicated.  At minimum, we have to include the magnetic
  sublattices.

\clearpage


\begin{table}
\caption{Computed effective spin angular momentum for each element in
  GdFeCo and TbFe alloys. Multiple pairs of alloys are used to compute
  the effective spin angular momentum for several compounds to
  demonstrate the range of the change in the spin angular
  momentum. The sign convention of the spin angular momentum is that
  either Gd or Tb has a positive value, while Fe has a negative
  value. The original signs of those spin angular momentum are shown
  in Tables \ref{tab1} through \ref{tab4}.  A simple cubic structure
  is adopted for Fe.}
\begin{tabular}{lllcc}
\hline
Alloy &$S^{eff}_{\rm Gd}(\hbar)$ &$S^{eff}_{\rm Fe}(\hbar)$ &
$S^{eff}_{\rm Tb}(\hbar)$ (orbital free) &$S^{eff}_{\rm Fe}(\hbar)$
(orbital free) \\
\hline
Gd$_{28}$Fe$_{63}$Co$_{9}$          & 1.3414   &  -1.1691  & -- &  --    \\
Gd$_{26}$Fe$_{64.7}$Co$_{9.3 }$     &   1.2456  &  -1.2006  & -- &  --   \\
Gd$_{25}$Fe$_{65.6}$Co$_{9.4}$      &  1.1517 &     -1.1777   & -- &--  \\
Gd$_{24}$Fe$_{66.5}$Co$_{9.5 }$     &  1.2262   &  -1.3017  & -- &  --   \\
Gd$_{24}$Fe$_{66.5}$Co$_{9.5 }$     &  1.0867   &    -1.0113   & -- &  --   \\
 Gd$_{22}$Fe$_{68.2}$Co$_{9.8}$     &   1.1241  &   -1.3350   & -- &  --  \\
Gd$_{22}$Fe$_{68.2}$Co$_{9.8}$      &   1.0135   &   -1.2244   & -- &  --
\\
Gd$_{22}$Fe$_{68.2}$Co$_{9.8}$      &   0.9846     &     -0.7737       & -- &  --
\\
Gd$_{22}$Fe$_{74.6}$Co$_{3.4}$      &     0.9846     &   -0.8463   & -- &  --   \\
\hline
Tb$_{30}$Fe$_{70}$ & -- &  --  &    2.1506  &    -1.8385 \\
Tb$_{30}$Fe$_{70}$   &-- & --   &       1.7594    &     -1.4473  \\
Tb$_{30}$Fe$_{70}$   &-- & --   &       1.4952        &      -1.1831
\\
Tb$_{30}$Fe$_{70}$   &-- & --   &          1.4698           &       -1.1577       \\
Tb$_{29}$Fe$_{71}$   &-- & --   &      2.0789      &   -1.8648    \\
Tb$_{27}$Fe$_{73}$  &-- & --  &       1.5835    &     -1.5093      \\
Tb$_{27}$Fe$_{73}$  &-- & --   &        1.1867    &     -1.1125     \\
Tb$_{24}$Fe$_{76}$  &-- &--   &       1.2641    &      -1.3452
\\
Tb$_{24}$Fe$_{76}$  &-- &--   &         1.1758         &      -1.2569         \\
Tb$_{22}$Fe$_{78}$   &-- & --   &     1.2789      &    -1.5007    \\
Tb$_{22}$Fe$_{78}$   &-- & --   &       1.1587           &    -1.3806
\\
Tb$_{22}$Fe$_{78}$   &-- & --   &          1.0965               &    -1.3183
\\
Tb$_{22}$Fe$_{78}$   &-- & --   &          0.9669 &        -1.1887
\\
\hline
\end{tabular}
\label{tab0}
\end{table}


\clearpage

\begin{table}
\caption{Computed effective spin angular momentum for each element in
  GdFeCo alloys for simple cubic Fe structure.}
\begin{tabular}{lrrr}
\hline
\hline
& & ~~$S^{eff}_{\rm Gd}$~~ &~~$S^{eff}_{\rm Fe}$~~ \\
\hline
 Gd&  -10.2187$\mu_{B}$&--&\\
 Fe&    3.9149$\mu_{B}$&--&\\
 Gd24Fe66.5Co9.5&&   -1.2262$\hbar$&    1.3017$\hbar$ \\ 
 Gd22Fe68.2Co9.8&&   -1.1241$\hbar$&    1.3350$\hbar$ \\ 
 \hline
 Gd&    9.5818$\mu_{B}$&--&\\
 Fe&   -3.7114$\mu_{B}$&--&\\
 Gd28Fe63Co9    &&    1.3414$\hbar$&   -1.1691$\hbar$ \\ 
 Gd26Fe64.7Co9.3&&    1.2456$\hbar$&   -1.2006$\hbar$ \\ 
 \hline
 Gd&   -9.2139$\mu_{B}$&--&\\
 Fe&    3.5907$\mu_{B}$&--&\\
 Gd25Fe65.6Co9.4&&   -1.1517$\hbar$&    1.1777$\hbar$ \\ 
 Gd22Fe68.2Co9.8&&   -1.0135$\hbar$&    1.2244$\hbar$ \\ 
 \hline
 Gd&   -9.0562$\mu_{B}$&--&\\
 Fe&    3.0415$\mu_{B}$&--&\\
 Gd24Fe66.5Co9.5&&   -1.0867$\hbar$&    1.0113$\hbar$ \\ 
 Gd22Fe74.6Co3.4&&   -0.9962$\hbar$&    1.1345$\hbar$ \\ 
 \hline
 Gd&    8.9509$\mu_{B}$&--&\\
 Fe&   -2.2689$\mu_{B}$&--&\\
 Gd22Fe68.2Co9.8&&    0.9846$\hbar$&   -0.7737$\hbar$ \\ 
 Gd22Fe74.6Co3.4&&    0.9846$\hbar$&   -0.8463$\hbar$ \\ 

\hline
\hline
\end{tabular}
\label{tab1}
\end{table}

\begin{table}
\caption{Computed effective spin angular momentum for each element in
  GdFeCo alloys for bcc Fe structure.}
\begin{tabular}{lrrr}
\hline
\hline
& & ~~$S^{eff}_{\rm Gd}$~~ &~~$S^{eff}_{\rm Fe}$~~ \\
\hline
 Gd&   -7.4560$\mu_{B}$&--&\\
 Fe&    2.8615$\mu_{B}$&--&\\
 Gd24Fe66.5Co9.5&&   -0.8947$\hbar$&    0.9514$\hbar$ \\ 
 Gd22Fe68.2Co9.8&&   -0.8202$\hbar$&    0.9758$\hbar$ \\ 
 \hline
 Gd&    7.4926$\mu_{B}$&--&\\
 Fe&   -2.9045$\mu_{B}$&--&\\
 Gd28Fe63Co9    &&    1.0490$\hbar$&   -0.9149$\hbar$ \\ 
 Gd26Fe64.7Co9.3&&    0.9740$\hbar$&   -0.9396$\hbar$ \\ 
 \hline
 Gd&   -6.7695$\mu_{B}$&--&\\
 Fe&    2.6400$\mu_{B}$&--&\\
 Gd25Fe65.6Co9.4&&   -0.8462$\hbar$&    0.8659$\hbar$ \\ 
 Gd22Fe68.2Co9.8&&   -0.7446$\hbar$&    0.9002$\hbar$ \\ 
 \hline
 Gd&   -6.6669$\mu_{B}$&--&\\
 Fe&    2.2355$\mu_{B}$&--&\\
 Gd24Fe66.5Co9.5&&   -0.8000$\hbar$&    0.7433$\hbar$ \\ 
 Gd22Fe74.6Co3.4&&   -0.7334$\hbar$&    0.8339$\hbar$ \\ 
 \hline
 Gd&    6.7532$\mu_{B}$&--&\\
 Fe&   -1.7221$\mu_{B}$&--&\\
 Gd22Fe68.2Co9.8&&    0.7428$\hbar$&   -0.5873$\hbar$ \\ 
 Gd22Fe74.6Co3.4&&    0.7428$\hbar$&   -0.6424$\hbar$ \\

\hline
\hline
\end{tabular}
\label{tab2}
\end{table}

\begin{table}
\caption{Computed effective spin angular momentum for each element in
TbFe alloys for simple cubic Fe structure.}
\begin{tabular}{lrrr}
\hline
\hline
& & ~~$S^{eff}_{\rm Tb}$~~ &~~$S^{eff}_{\rm Fe}$~~ \\
\hline

 Tb&   14.3375$\mu_{B}$&--&\\
 Fe&   -5.2529$\mu_{B}$&--&\\
 Tb30Fe70&&    2.1506$\hbar$&   -1.8385$\hbar$ \\ 
 Tb29Fe71&&    2.0789$\hbar$&   -1.8648$\hbar$ \\ 
 \hline
 Tb&   11.7296$\mu_{B}$&--&\\
 Fe&   -4.1352$\mu_{B}$&--&\\
 Tb30Fe70&&    1.7594$\hbar$&   -1.4473$\hbar$ \\ 
 Tb27Fe73&&    1.5835$\hbar$&   -1.5093$\hbar$ \\ 
 \hline
 Tb&  -11.6262$\mu_{B}$&--&\\
 Fe&    3.8479$\mu_{B}$&--&\\
 Tb22Fe78&&   -1.2789$\hbar$&    1.5007$\hbar$ \\ 
 Tb19Fe81&&   -1.1045$\hbar$&    1.5584$\hbar$ \\ 
 \hline
 Tb&  -10.5340$\mu_{B}$&--&\\
 Fe&    3.5399$\mu_{B}$&--&\\
 Tb22Fe78&&   -1.1587$\hbar$&    1.3806$\hbar$ \\ 
 Tb24Fe76&&   -1.2641$\hbar$&    1.3452$\hbar$ \\ 
 \hline
 Tb&    9.9679$\mu_{B}$&--&\\
 Fe&   -3.3802$\mu_{B}$&--&\\
 Tb22Fe78&&    1.0965$\hbar$&   -1.3183$\hbar$ \\ 
 Tb30Fe70&&    1.4952$\hbar$&   -1.1831$\hbar$ \\ 
 \hline
 Tb&    9.7986$\mu_{B}$&--&\\
 Fe&   -3.3076$\mu_{B}$&--&\\
 Tb30Fe70&&    1.4698$\hbar$&   -1.1577$\hbar$ \\ 
 Tb24Fe76&&    1.1758$\hbar$&   -1.2569$\hbar$ \\ 
 \hline
 Tb&   -8.7901$\mu_{B}$&--&\\
 Fe&    3.0480$\mu_{B}$&--&\\
 Tb22Fe78&&   -0.9669$\hbar$&    1.1887$\hbar$ \\ 
 Tb27Fe73&&   -1.1867$\hbar$&    1.1125$\hbar$ \\ 
 \hline
 Tb&    5.3268$\mu_{B}$&--&\\
 Fe&   -1.3912$\mu_{B}$&--&\\
 Tb36Fe64&&    0.9588$\hbar$&   -0.4452$\hbar$ \\ 
 Tb30Fe70&&    0.7990$\hbar$&   -0.4869$\hbar$ \\ 
 \hline
 Tb&    2.1818$\mu_{B}$&--&\\
 Fe&   -0.4756$\mu_{B}$&--&\\
 Tb34Fe66&&    0.3709$\hbar$&   -0.1570$\hbar$ \\ 
 Tb24Fe76&&    0.2618$\hbar$&   -0.1807$\hbar$ \\

\hline
\hline
\end{tabular}
\label{tab3}
\end{table}

\begin{table}
\caption{Computed effective spin angular momentum for each element in
TbFe alloys for bcc Fe structure.}
\begin{tabular}{lrrr}
\hline
\hline
& & ~~$S^{eff}_{\rm Tb}$~~ &~~$S^{eff}_{\rm Fe}$~~ \\
\hline
 Tb&   11.2034$\mu_{B}$&--&\\
 Fe&   -4.1158$\mu_{B}$&--&\\
 Tb30Fe70&&    1.6805$\hbar$&   -1.4405$\hbar$ \\ 
 Tb29Fe71&&    1.6245$\hbar$&   -1.4611$\hbar$ \\ 
 \hline
 Tb&    9.0822$\mu_{B}$&--&\\
 Fe&   -3.2067$\mu_{B}$&--&\\
 Tb30Fe70&&    1.3623$\hbar$&   -1.1224$\hbar$ \\ 
 Tb27Fe73&&    1.2261$\hbar$&   -1.1705$\hbar$ \\ 
 \hline
 Tb&   -7.8075$\mu_{B}$&--&\\
 Fe&    2.6098$\mu_{B}$&--&\\
 Tb22Fe78&&   -0.8588$\hbar$&    1.0178$\hbar$ \\ 
 Tb19Fe81&&   -0.7417$\hbar$&    1.0570$\hbar$ \\ 
 \hline
 Tb&   -7.4627$\mu_{B}$&--&\\
 Fe&    2.5126$\mu_{B}$&--&\\
 Tb22Fe78&&   -0.8209$\hbar$&    0.9799$\hbar$ \\ 
 Tb24Fe76&&   -0.8955$\hbar$&    0.9548$\hbar$ \\ 
 \hline
 Tb&    7.4621$\mu_{B}$&--&\\
 Fe&   -2.5124$\mu_{B}$&--&\\
 Tb22Fe78&&    0.8208$\hbar$&   -0.9798$\hbar$ \\ 
 Tb30Fe70&&    1.1193$\hbar$&   -0.8793$\hbar$ \\ 
 \hline
 Tb&    7.4619$\mu_{B}$&--&\\
 Fe&   -2.5123$\mu_{B}$&--&\\
 Tb30Fe70&&    1.1193$\hbar$&   -0.8793$\hbar$ \\ 
 Tb24Fe76&&    0.8954$\hbar$&   -0.9547$\hbar$ \\ 
 \hline
 Tb&   -6.3789$\mu_{B}$&--&\\
 Fe&    2.2069$\mu_{B}$&--&\\
 Tb22Fe78&&   -0.7017$\hbar$&    0.8607$\hbar$ \\ 
 Tb27Fe73&&   -0.8611$\hbar$&    0.8055$\hbar$ \\ 
 \hline
 Tb&    4.4942$\mu_{B}$&--&\\
 Fe&   -1.2404$\mu_{B}$&--&\\
 Tb36Fe64&&    0.8090$\hbar$&   -0.3969$\hbar$ \\ 
 Tb30Fe70&&    0.6741$\hbar$&   -0.4342$\hbar$ \\ 
 \hline
 Tb&    1.7919$\mu_{B}$&--&\\
 Fe&   -0.4100$\mu_{B}$&--&\\
 Tb34Fe66&&    0.3046$\hbar$&   -0.1353$\hbar$ \\ 
 Tb24Fe76&&    0.2150$\hbar$&   -0.1558$\hbar$ \\ 

\hline\hline
\end{tabular}
\label{tab4}
\end{table}

\clearpage

\end{document}